\PassOptionsToPackage{prologue,dvipsnames}{xcolor}
\documentclass[twocolumn]{aastex631}

\usepackage{showyourwork}
\usepackage{bm}
\usepackage{amsmath}
\usepackage{booktabs}
\usepackage{appendix}
\usepackage{ifthen}
\usepackage{array}
\usepackage{soul}
\usepackage{float}
\input{macros.tex}
\usepackage{multirow}
\newcommand{\CIPlusMinus}[1]{{#1[median]^{+#1[error plus]}_{-#1[error minus]}}}
\newcommand{\CIPlusMinusPer}[1]{{#1[median]\%^{+#1[error plus]\%}_{-#1[error minus]\%}}}

\newcommand{\msun}{\ensuremath{{\,\rm M}_\odot}}
\newcommand{\popA}{\textsc{SpinPop\textsubscript{A}}}
\newcommand{\popB}{\textsc{SpinPop\textsubscript{B}}}
\newcommand{\first}{\popA{}\textsc{:Peak}}
\newcommand{\contB}{\popB{}\textsc{:Continuum}}
\newcommand{\contA}{\popA{}\textsc{:Continuum}}
\newcommand{\brucepaper}{\citet{2022arXiv221012834E}}
\newcommand{\othreea}{\citet{2010.14533}}
\newcommand{\cont}[1]{\textcolor{red}{DELTE THIS}}
\newcommand{\second}[1]{\textcolor{red}{DELTE THIS}}
\newcommand{\base}{\textsc{Isolated Peak Model}}
\newcommand{\comp}{\textsc{Peak+Continuum Model}}

\newcommand{\result}[1]{\textcolor{BurntOrange}{#1}}

\graphicspath{{./}{figures/}}

\begin{document}
\title{Cosmic Cousins: Identification of a Subpopulation of Binary Black Holes Consistent with Isolated Binary Evolution}

\author{Jaxen Godfrey}
\email{jaxeng@uoregon.edu}
\affiliation{Institute  for  Fundamental  Science, Department of Physics, University of Oregon, Eugene, OR 97403, USA}
\author{Bruce Edelman}
\affiliation{Institute  for  Fundamental  Science, Department of Physics, University of Oregon, Eugene, OR 97403, USA}
\author{Ben Farr}
\affiliation{Institute  for  Fundamental  Science, Department of Physics, University of Oregon, Eugene, OR 97403, USA}

\begin{abstract}
    Observations of gravitational waves (GWs) from merging compact binaries have become a regular occurrence. The continued advancement of the LIGO-Virgo-KAGRA (LVK) detectors have now produced a catalog of over 90 such mergers, from which we can begin to uncover the formation history of merging compact binaries. In this work, we search for subpopulations in the LVK's third gravitational wave transient catalog through the use of data-driven mixture models in a hierarchal Bayesian inference framework. By allowing for unique correlations between mass and spin in each subpopulation, we find an over density of mergers with a primary mass of $\sim10\msun$, consistent with isolated binary formation. This low-mass subpopulation has a spin magnitude distribution peaking at $a_\mathrm{peak}=$ \result{$\CIPlusMinus{\macros[SpinMag][Base][PeakA][max]}$}, exhibits spins preferentially aligned with the binary's orbital angular momentum, is constrained by \result{$\CIPlusMinus{\macros[NumEvents][Base][PeakA]}$} of our observations, and contributes \result{$\CIPlusMinusPer{\macros[BranchingRatios][Base][PeakA][Percent]}$} to the overall population of BBHs. We find a $\sim\macros[FracCut][BgreaterA]\%$ chance that the bulk of the events in the $15-55\msun$ range share a spin distribution with the $10\msun$ peak, with $99\%$ of these events possessing primary masses less than $m_{1,99\%} = $ \result{$\CIPlusMinus{\macros[Mass][Composite][ContinuumA][99percentile]}\msun$}, providing an estimate of the lower edge of the theorized pair instability mass gap. Additionally, we find mild evidence for a subpopulation of high mass BBHs near $60\msun$. This work is a first step in gaining a deeper understanding of compact binary formation and evolution with data-driven models, and will provide more robust conclusions as the catalog of observations becomes larger. 
\end{abstract}

\section{Introduction} \label{sec:intro}

The first detection of gravitational waves (GWs) from a binary black hole (BBH) merger was made by the LIGO-Virgo-KAGRA (LVK) Collaboration on September 14, 2015. Since that fateful day, the LVK has detected nearly 100 compact binary coalescences (CBCs), bringing the third gravitational wave transient catalog (GWTC-3) up to 90 such events. \citep{2015CQGra..32g4001L,2015CQGra..32b4001A, 2016PhRvL.116f1102A,2019PhRvX...9c1040A,2021PhRvX..11b1053A,2021arXiv211103606T, 2021PTEP.2021eA102A}. With the maturation of GW Astronomy, novel studies of the universe are possible; we are now able to probe the low-redshift population of merging compact objects with much greater fidelity than with the sparse, early LVK catalogs \citep{2019ApJ...882L..24A,2021ApJ...913L...7A,2021arXiv211103634T}. By breaking down the full CBC population into subpopulations based on different source properties -- and paired with our theoretical knowledge of stellar astrophysics -- we can begin to uncover the formation and evolution of compact binaries \citep{10.1088/0004-637X/810/1/58, 2017ApJ...846...82Z}. While there are many proposed processes that could lead to a compact binary able to merge through gravitational radiation, they largely fall within two broad categories: isolated formation and dynamical assembly.  Each of these channels could leave their own distinct imprint on the binaries they produce, including the distributions of masses and spins \citep{2017Natur.548..426F,2018ApJ...854L...9F,10.3847/1538-4357/ab88b2}. The uncertainty in modeled merger rates of each formation channel is large, and the predictions continue to evolve with better understanding of the underlying physics (see \citet{10.1007/s41114-021-00034-3} for a thorough review on both modeled and observed merger rates of compact objects). By looking deeper at the correlations between the source properties at a population level, the identification of distinct subpopulations with common mass or spin properties could help to identify unique formation mechanisms or channels.

Isolated formation of compact binaries occurs in galactic fields where two progenitors are born gravitationally bound, isolated from interactions with other stars, and undergo standard main sequence evolution, each eventually forming into a compact object. Energy loss from GW radiation causes the binary to inspiral, which can eventually lead to merger; however, in order for the binary to merge within a Hubble time, the initial orbital separation of the compact objects must be small \citep{10.1051/0004-6361/201936204,10.1007/s41114-021-00034-3}. For the binary to survive the early evolutionary phases of the progenitors, the orbit must initially be wide; thus, some process during late-stage evolution is required to significantly reduce the orbital separation before the second compact object is formed. One proposed mechanism is the ``common envelope phase'', where both stars momentarily share an envelope (typically after one star has already collapsed into a compact object) and drag forces quickly dissipate orbital energy, which reduces the orbital separation enough so that the resulting compact binary can merge due to GW emission alone \citep{10.1038/nature18322}. In dynamical formation scenarios, scattering and exchange interactions between astrophysical bodies in a dense stellar environment are thought to produce binaries capable of merging within a Hubble time \citep{1602.02444}. There are many theorized models for the main physical processes that contribute to isolated and dynamical formation, but there are few robust and direct predictions of observable quantities from these models. Instead, current predictions of merger rates and population distributions are estimated from simulations, which have large uncertainties due to uncertain underlying physics or poorly constrained initial conditions \citep{1308.1546, 1806.00001v3, 10.1051/0004-6361/201936204, 10.1007/s41114-021-00034-3}.

The spin distributions of merging binaries are thought to provide the clearest distinction between possible formation channels \citep{10.1088/1361-6382/aa552e, 10.1093/mnras/stx1764, 2017Natur.548..426F,2018ApJ...854L...9F}. Systems assembled dynamically in stellar clusters are expected to have no preferential alignment, producing an isotropically distributed spin tilt distribution \citep{1609.05916,10.1103/PhysRevD.100.043027}, though there are no robust predictions of spin magnitude from this channel. Binaries formed in the field can be born with a range of spin magnitude values -- dependendent upon the efficiency of angular momentum transport and other physical processes -- that are predicted to be aligned with the orbital angular momentum of the binary. With current data it is difficult to distinguish between these two channels, though studies have at least shown that GWTC-3 is not consistent with entirely dynamical or entirely isolated formation \citep{2021arXiv211103634T,2011.09570,2022ApJ...937L..13C,2022arXiv220902206T,2022arXiv221012834E,10.3847/2041-8213/ac86c4}. Recent studies have found support for a significant contribution of systems formed through dynamical assembly in the population of BBHs inferred from the GWTC-2 and GWTC-3 catalogs \citep{2021ApJ...913L...7A,2021ApJ...921L..15G, 2021PhRvD.104h3010R,2021arXiv211103634T,2022ApJ...937L..13C,2022arXiv220902206T,2022arXiv220906978V,2022arXiv221012834E}, though with large uncertainties.

While spin may be the clearest indicator of compact binary formation history, spin measurements of individual binaries typically have large uncertainties, making it difficult to disentangle competing formation channels with spin alone. However, the component masses of individual events are typically inferred with higher precision than their spins, and there are even features in the mass distribution that may signal the existence of different subpopulations \citep{2021ApJ...913L..19T, 2021arXiv211103634T, 2022ApJ...928..155T, 2022ApJ...924..101E,2022arXiv221012834E, 2208.11871, 2209.05766, 2301.00834}. Unfortunately, it can also be challenging to distinguish between the isolated and dynamical formation channels using only component mass, as the models in both scenarios predict masses that significantly overlap \citep{1609.05916}. Instead, a search for correlated population properties across mass, spin, and redshift may prove to be much more fruitful in distinguishing between the different compact binary formation channels \citep{2021ApJ...912...98F,2021ApJ...922L...5C,2022ApJ...931...17V,2022ApJ...932L..19B}.

Given the evidence for features, e.g. peaks or truncations, in the primary mass spectrum \citep{10.3847/2041-8213/aa9bf6, 10.3847/1538-4357/aab34c, 10.3847/2041-8213/ab3800, 2021ApJ...913L...7A, 2111.03634,2104.07783v2, 2022ApJ...924..101E, 2022arXiv221012834E,2022ApJ...928..155T, 2301.00834,10.48550/arXiv.2302.07289}, in this work we attempt to identify subpopulations of BBHs grouped in primary mass with spin properties distinct from the rest of the GWTC-3 BBH population. We do this by modeling a portion of the BBH primary mass spectrum as a Gaussian peak and allowing the spin distributions of the events categorized in this peak to differ from the rest of the population. In this work, we present strong evidence for a distinct subpopulation at $10 \msun$, and conduct a follow-up search for binaries elsewhere in the mass spectrum with similar spin properties.

The remaining sections of this manuscript are structured as follows: Section \ref{sec:methods} describes the statistical framework and the different component models studied. Section \ref{sec:results} presents the results of our study, including the inferred branching ratios and the inferred subpopulation membership probabilities for each BBH in GWTC-3. In Section \ref{sec:astro} we discuss the implications of our findings and how it relates to the current understanding of compact binary formation and population synthesis. We finish in section \ref{sec:conclusion}, with a summary of the letter and prospects for distinguishing subpopulations in future catalogs after the LVK's fourth observing run.
\section{Methods} \label{sec:methods}

\subsection{Statistical Framework} \label{sec:statistical_framework}

We employ the typical hierarchical Bayesian inference framework to infer the properties of the population of merging compact binaries given a catalog of observations. The rate of compact binary mergers is modeled as an inhomogeneous Poisson point process \citep{10.1093/mnras/stz896}, with the merger rate per comoving volume $V_c$ \citep{astro-ph/9905116}, source-frame time $t_\text{src}$ and binary parameters $\theta$ defined as:

\begin{equation} \label{eq:rate}
    \mathcal{R} = \frac{dN}{dV_cdt_\mathrm{src}d\theta} = \frac{dN}{dV_cdt_\mathrm{src}} p(\theta | \Lambda)
\end{equation}

\noindent with $p(\theta | \Lambda)$ the population model, $\mathcal{R}$ the merger rate, and $\Lambda$ the set of population hyperparameters. Following other population studies \citep{10.1093/mnras/stz896,2021ApJ...913L...7A,2111.03634,2007.05579}, we use the hierarchical likelihood \citep{10.1063/1.1835214} that incorporates selection effects and marginalizes over the merger rate as: 

\begin{equation} \label{eq:likelihood}
    \mathcal{L}(\bm{d} | \Lambda) \propto \frac{1}{\xi(\Lambda)} \prod_{i=1}^{N_\mathrm{det}} \int d\theta \mathcal{L}(d_i | \theta) p(\theta | \Lambda)
\end{equation}

\noindent Above, $\bm{d}$ is the set of data containing $N_\mathrm{det}$ observed events, $\mathcal{L}(d_i | \theta)$ is the individual event likelihood function for the $i$th event given parameters $\theta$ and $\xi(\Lambda)$ is the fraction of merging binaries we expect to detect, given a population described by $\Lambda$. The integral of the individual event likelihoods marginalizes over the uncertainty in each event's binary parameter estimation, and is calculated with Monte Carlo integration and by importance sampling, reweighing each set of posterior samples to the likelihood. The detection fraction is calculated with:

\begin{equation} \label{eq:detfrac}
    \xi(\Lambda) = \int d\theta p_\mathrm{det}(\theta) p(\theta | \Lambda)
\end{equation}

\noindent with $p_\mathrm{det}(\theta)$ the probability of detecting a binary merger with parameters $\theta$. We calculate this fraction using simulated compact merger signals that were evaluated with the same search algorithms that produced the catalog of observations. With the signals that were successfully detected, we again use Monte Carlo integration to get the overall detection efficiency, $\xi(\Lambda)$ \citep{1712.00482, 1904.10879, 2204.00461}.



\begin{equation} \label{eq:latent}
    p(\theta_i | \Lambda, k_i) = p_{k_i}(\theta_i | \Lambda_{k_i})
\end{equation}


\subsection{Astrophysical Mixture Models} \label{sec:astromodels} 

Given the recent evidence for peaks \citep{10.3847/2041-8213/ab3800, 2021ApJ...913L...7A, 2111.03634, 2022ApJ...928..155T,2022ApJ...924..101E, 2022arXiv221012834E,10.48550/arXiv.2302.07289} in the BBH primary mass spectrum at $10 M_{\odot}$, $35 M_{\odot}$, and suggestions of a potential feature at $\sim20 M_{\odot}$, we chose mixture models similar to the \textsc{Multi Spin} model in \cite{2021ApJ...913L...7A}, which is characterized by a power-law plus a Gaussian peak in primary mass, wherein the spin distributions of the two components in mass are allowed to differ from each other. In our case, we replace the power law mass components and parametric spin descriptions with fully data-driven (often called "non-parametric") Basis-Spline functions. 

We make use of the mass and spin Basis-Spline (B-Spline) models from \cite{2022arXiv221012834E}, with a few modifications. We fix the number of knots $n$ in all B-Spline models used, choosing $n_{m_1}=48$, $n_{q}=30$, $n_a=16$, and $n_{cos(\theta)}=16$. When used in a Bayesian inference setting, a prior can be used to place a smoothing requirement on the B-Spline function by penalizing large differences between neighboring coefficients. This penalization allows the user to choose a large number of basis functions (i.e. coefficients) to accurrately fit the data without concern for overfitting. Our coefficient and smoothing priors have the same form as in equations B4 and B5 in \brucepaper{}, though we fix the value of the smoothing prior scale $\tau_\lambda$ for each B-Spline model. we found that $\tau_\lambda$ consistently railed against boundaries set by the minimum required effective sample size in single-event Monte Carlo integrals, wherein the minimum effective sample size was the main driver of smoothness. We therefore opted to fix $\tau_\lambda$ for each population distribution to reasonable values that produced results consistent with other non-parametric results like \citet{2022ApJ...924..101E}, \brucepaper{}, and \citet{10.48550/arXiv.2302.07289}. We plan to address this issue in a future work by exploring different coefficient and smoothing prior specifications, such as locally adaptive smoothing prior. For a full description of the priors used in each model, see Appendix \ref{sec:priors}.

In both of our model prescriptions, detailed below, the spin magnitude and tilt distributions of each binary component are assumed to be independently and identically distributed (IID); i.e. both binary components have spins drawn from the same (inferred) distribution. In both cases we'll describe the overall population as a two-component mixture model, with each component (which we refer to as \popA{} and \popB{}) having a unique spin distribution. We do not fit the mininum and maximum BBH primary mass, instead truncating all the mass distributions below $m_\text{min} = 5.0\msun$ and above $m_\text{max} = 100\msun$; however, estimates of minimum and maximum BBH mass are still possible as B-Splines are able to become arbitarily small in locations where the data requires it, effectively placing upper bounds on merger rate densities in regions without detections, consistent with the observed $VT$. Finally, we estimate the redshift distribution with a modulated powerlaw distribution as in \brucepaper, and assume it is the same for each subpopulation:

\begin{equation}
    p(z|\Lambda_z) = \frac{dV}{dz} (1+z)^{\lambda - 1} \exp \mathcal{B}(\text{log}z).
\end{equation}

To construct our probabilistic mixture model, we sample $p_{\lambda} \sim \mathcal{D}(M)$, from an M-dimensional Dirichlet distribution of equal weights, representing the astrophysical branching ratios $\lambda_{i}$ of each subpopulation. The total population model $p(\theta | \{\Lambda\})$ is then a weighted sum of each subpopulation model $p_i(\theta|\Lambda_i)$ with $\lambda_i$ as the weights:

\begin{equation} \label{totmixmod}
p(\theta|\Lambda) = \sum_{i=0}^{M} \lambda_i p_i(\theta | \Lambda_i)
\end{equation}

Within the \textsc{NumPyro} \citep{1810.09538,1912.11554} probabilistic programming language, we use the \texttt{NUTS} \citep{1111.4246} sampler to perform our parameter estimation. All the models and formalism used in our analysis are available in the \href{https://github.com/FarrOutLab/GWInferno}{GWInferno} python library, along with the code and data to reproduce this study in this GitHub \href{https://github.com/jaxeng/CosmicCousins}{repository}.

\subsubsection{\base{}}
In this model prescription, we categorize the BBH population into $M=2$ subpopulations, which we will refer to as \popA{} and \popB{}. \popA is characterised by a truncated Gaussian peak in log-primary mass and B-Spline functions in mass ratio, spin magnitude, and tilt angle. \popB{} is characterized by B-Spline functions in all mass and spin parameters. We infer the mean $\mu_m$ and standard deviation $\sigma_m$ of the peak, along with the B-Spline coefficients $\mathbf{c}$ for all other parameters.

\begin{itemize}
    \item \popA{}, $i=0$. This category assumes a truncated Gaussian model ($\mathcal{N}_\text{T}$) in log-primary mass and B-spline models ($\mathcal{B}$) in mass ratio $q$, spin magnitude $a_j$, and tilt $cos(t_j)$. Note that since we assume the spins to be IID, the B-Spline spin parameters, such as $\mathbf{c}_{a,0}$, are the same for each binary component $j=1$ and $j=2$
    \begin{eqnarray} \label{eq:peakAbase}
        \text{log} p_{m,0}(\text{log} m_1| \Lambda_{m,0}) = \mathcal{N}_\text{T}(\text{log}m_1 | \mu_{m}, \sigma_{m}) \\
        \text{log} p_{q,0}(q| \Lambda_{q,0}) = \mathcal{B}(q | \mathbf{c}_{q,0}) \\
        \text{log} p_{a,0}(a_j| \Lambda_{a,0}) = \mathcal{B}(a_j | \mathbf{c}_{a,0}) \\
        \text{log} p_{t,0}(cos(t_j)| \Lambda_{t,0}) = \mathcal{B}( cos(t_j) | \mathbf{c}_{t,0})
    \end{eqnarray}

    \item \popB{}, $i=1$. The mass ratio and spin models have the same form as \popA{}, but here primary mass is fit to a B-spline function. 
    \begin{eqnarray} \label{eq:contBbase}
        \text{log} p_{m,1}(m_1| \Lambda_{m,1}) = \mathcal{B}( \text{log} m_1 | \mathbf{c}_{m}) \\
        \text{log} p_{q,1}(q| \Lambda_{q,0}) = \mathcal{B}(q | \mathbf{c}_{q,1}) \\
        \text{log} p_{a,1}(a_j| \Lambda_{a,1}) = \mathcal{B}(a_j | \mathbf{c}_{a,1}) \\
        \text{log} p_{t,1}(cos(t_j)| \Lambda_{t,1}) = \mathcal{B}( cos(t_j) | \mathbf{c}_{t,1})
    \end{eqnarray}

\end{itemize}

\subsubsection{\comp{}}

To investigate if other parts of the primary mass spectrum have similar spin characteristics to \popA{} of the \base{}, we construct a composite subpopulation model whose primary mass is itself a mixture model of a Gaussian peak and a B-Spline, which we'll refer to as \first{} and \contA{}, respectively.


\begin{itemize}
    \item \popA{}, $i=0$ for \first{}, $i=1$ for \contA{}. This category assumes a truncated Gaussian model in primary mass for the \first{} mass component and a B-spline in primary mass for the \contA{} mass component. Both mass components are also fit to their own B-spline mass ratio models. Spin magnitude $a_j$, and $cos(t_{j})$ are described by B-splines. \textit{Note: the branching fraction for \contA{} is $\lambda_1$, which is allowed to differ from $\lambda_0$, the branching fraction for \first{}.}
    \begin{eqnarray} \label{eq:peakAcomp}
        \text{log} p_{m,0}(\text{log} m_1| \Lambda_{m,0}) = \mathcal{N}_\text{T}(\text{log}m_1 | \mu_{m}, \sigma_{m}) \\
        \text{log} p_{m,1}(\text{log} m_1| \Lambda_{m,1}) = \mathcal{B}(\text{log} m_1 | \mathbf{c}_{m, 1}) \\
        \text{log} p_{q,0}(q| \Lambda_{q,0}) = \mathcal{B}(q | \mathbf{c}_{q,0}) \\
        \text{log} p_{q,1}(q| \Lambda_{q,1}) = \mathcal{B}(q | \mathbf{c}_{q,1}) \\
        \text{log} p_{a,0}(a_j| \Lambda_{a,0}) = \mathcal{B}(a_j | \mathbf{c}_{a,0}) \\
        \text{log} p_{t,0}(cos(t_j)| \Lambda_{ta,0}) = \mathcal{B}( cos(t_j) | \mathbf{c}_{t,0})
    \end{eqnarray}

    \item \popB{}, $i=2$. Here, all mass and spin properties are fit to B-Splines.
    \begin{eqnarray} \label{eq:contBcomp}
        \text{log} p_{m,2}(\text{log}m_1| \Lambda_{m,2}) = \mathcal{B}(\text{log}m_1 | \mathbf{c}_{m, 2}) \\
        \text{log} p_{q,2}(q| \Lambda_{q,2}) = \mathcal{B}(q | \mathbf{c}_{q,2}) \\
        \text{log} p_{a,2}(a_j| \Lambda_{a,2}) = \mathcal{B}(a_j | \mathbf{c}_{a,2}) \\
        \text{log} p_{t,2}(cos(t_j)| \Lambda_{t,2}) = \mathcal{B}( cos(t_j) | \mathbf{c}_{t,2})
    \end{eqnarray}
\end{itemize}
\section{Results} \label{sec:results}

\begin{figure*}[ht!]
    \begin{centering}
        \includegraphics[width=\linewidth]{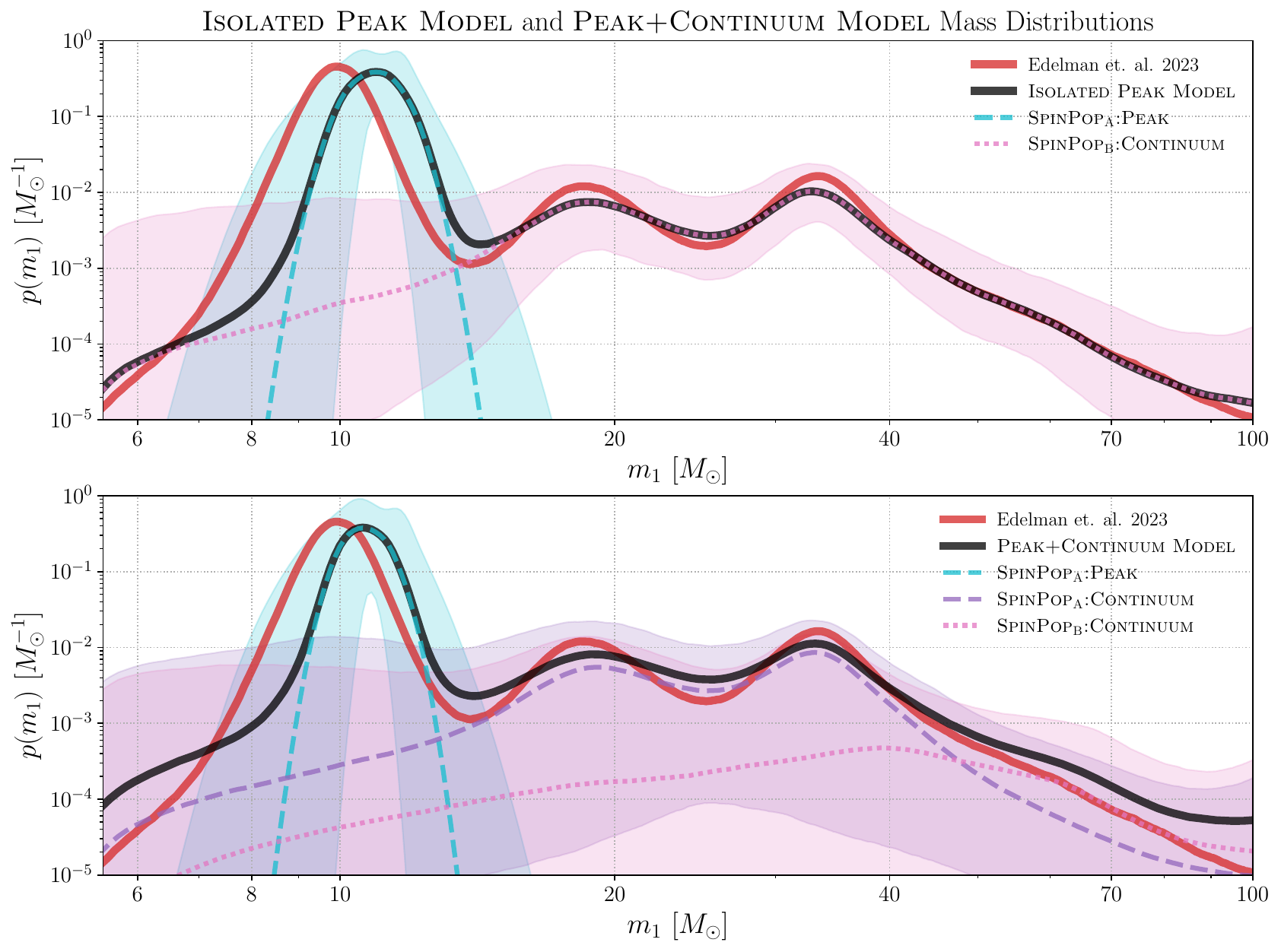}
        \caption{The astrophysical primary mass distribution inferred by the \base{} (top) and the \comp{} (bottom left). In both panels, the median distribution inferred by \brucepaper is shown in red and the total inferred by this work is shown in black. The median of the subpopulation component distributions are shown in dashed lines and the shaded regions indicate the $90\%$ credible regions. The subpopulation distributions are weighted by their respective astrophysical branching ratios.}
        \label{fig:g1_mass_distribution}
    \end{centering}
    \script{mass_distribution_g1_plot.py}
\end{figure*}

With these models and framework in hand, we infer the mass, spin and redshift distributions with the most recent LVK catalog of gravitational wave observations, GWTC-3 \citep{2021arXiv211103606T}. We perform the same BBH threshold cuts on the catalog done by the LVK's accompanying population analysis \othreea{}. For the remaining 69 BBH mergers, we follow what was done in \citet{2021arXiv211103606T}: for events included in GWTC-1 \citep{2019ApJ...882L..24A}, we use the published samples that equally weight samples from analyses with the \textsc{IMRPhenomPv2} \citep{1308.3271} and \textsc{SEOBNRv3} \citep{1307.6232,1311.2544} waveforms; for the events from GWTC-2 \citep{2021ApJ...913L...7A}, we use samples that equally weight all available analyses using higher-order mode waveforms (\textsc{PrecessingIMRPHM}); finally, for new events reported in GWTC-2.1 and GWTC-3 \citep{2021arXiv211103606T,2108.01045}, we use combined samples, equally weighted, from analyses with the \textsc{IMRPhenomXPHM} \citep{2004.06503} and the \textsc{SEOBNRv4PHM} \citep{2004.09442} waveform models.

\begin{figure}[ht!]
    \begin{centering}
        \includegraphics[width=\linewidth]{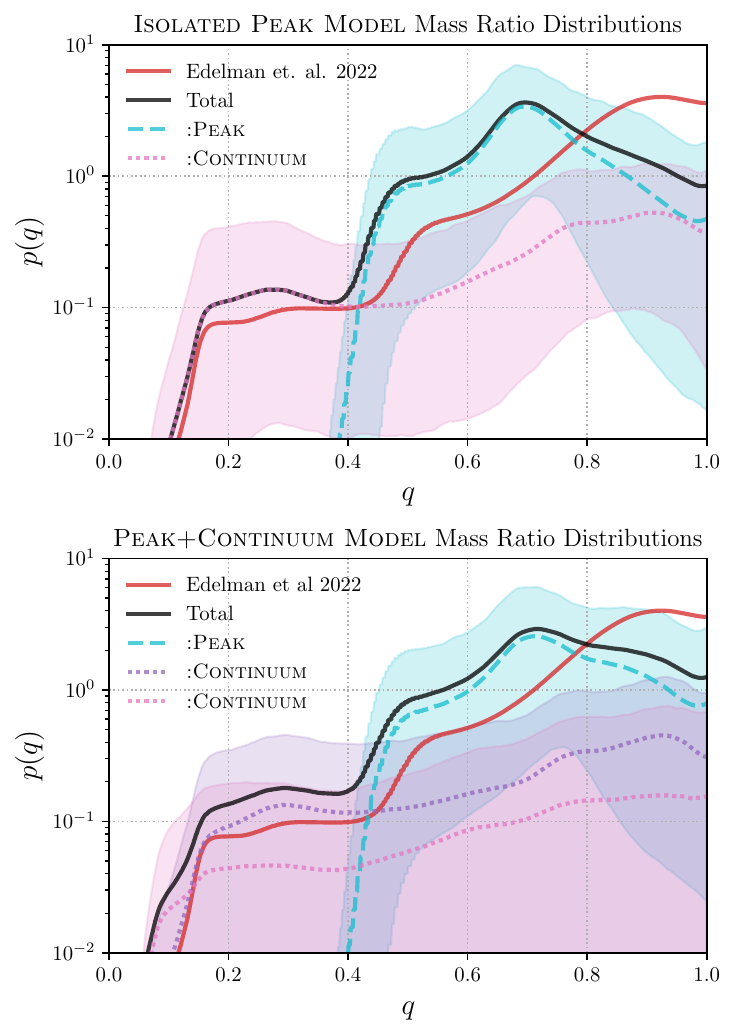}
        \caption{The astrophysical mass ratio distributions inferred by the \base{} (top) and the \comp{} (bottom). In both panels, the median distribution inferred by \brucepaper is shown in red and the total inferred by this work is shown in black. The median of the subpopulation component distributions are shown in dashed lines and the shaded regions indicate the $90\%$ credible regions. The subpopulation distributions are weighted by their respective branching ratios.}
        \label{fig:mass_ratio_distribution}
    \end{centering}
    \script{mass_ratio_distribution_plot.py}
\end{figure}

\subsection{BBH Mass and Spin Distributions}

Figure \ref{fig:g1_mass_distribution} shows the primary mass distributions of the \base{}, as well as the total primary mass distribution inferred by \brucepaper{} for comparison. As seen in Figure \ref{fig:g1_mass_distribution}, the total BBH primary mass distribution of the \base{} is statistically consistent with that inferred by \brucepaper{}. Inspecting the subpopulations, \first{} and \contB{}, we see that \first{} identifies a peak in the primary mass spectrum near $10\msun$ while \contB{} describes the rest of the spectrum above the peak. Interestingly, even though the $35\msun$ peak was the first observed departure from power law-like behavior in the mass spectrum, the $10\msun$ peak is the dominant feature that is isolated by the \comp{} and the spin properties of this subpopulation appear distinct from the broader catalog.

Figure \ref{fig:mass_ratio_distribution} shows the mass ratio distributions of the \base{} and \comp{}. Though uncertainties are large, the distributions across subpopulations have consistent features (the sharp fall-off of \first{} near $q=0.4$ in both models is due to the fixed minimum mass $m_\text{min}=5\msun$, and events in \first{} are $\lesssim10\msun$). The total mass ratio distributions for both models are consistent with the total inferred by \brucepaper.

\begin{figure*}[]
    \begin{centering}
        \includegraphics[width=\linewidth]{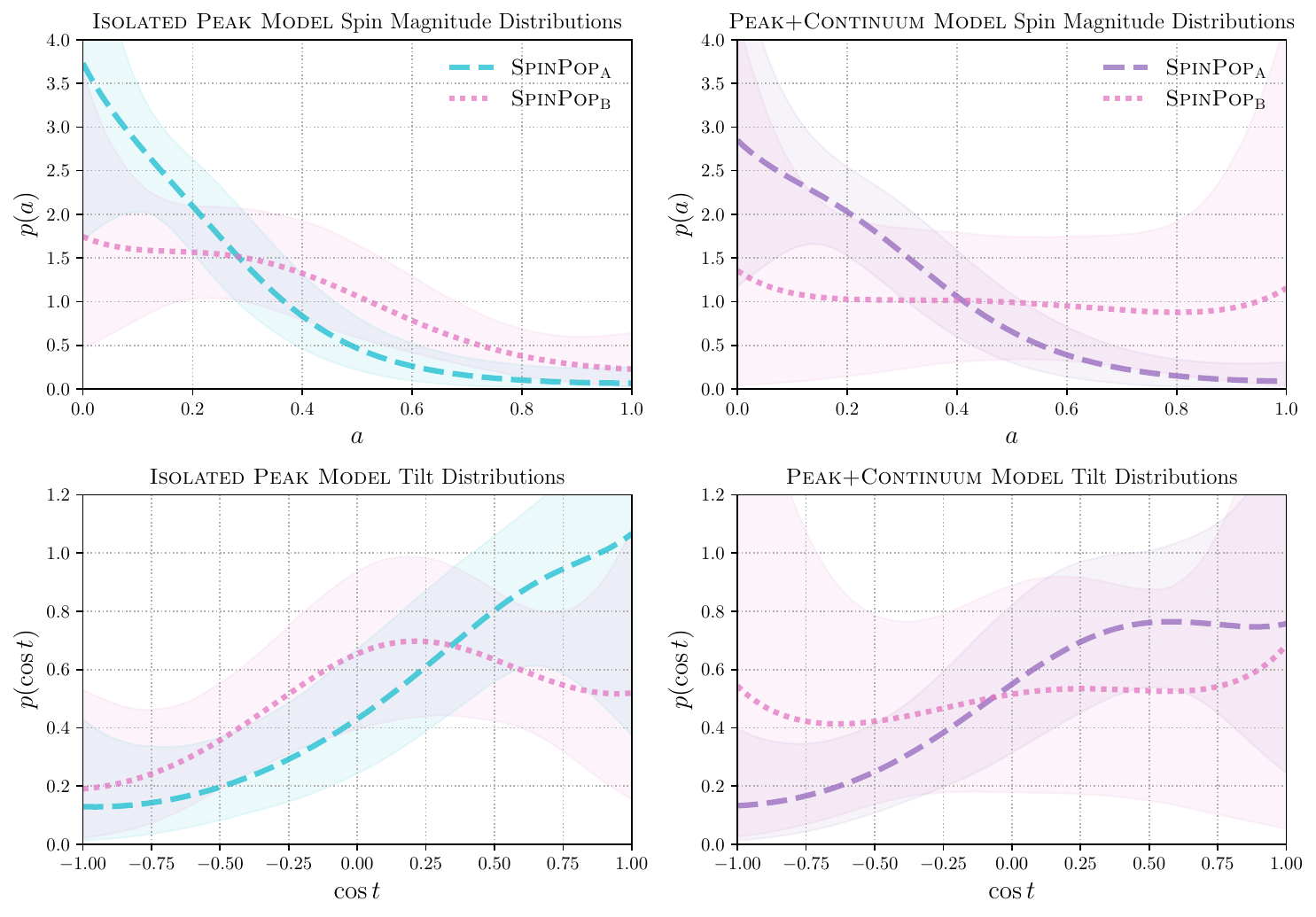}
        \caption{The spin magnitude and tilt distributions inferred by the \base{} (top left and bottom left) and the \comp{} (top right and bottom right). In each panel, the medians of the subpopulation component distributions are shown in dashed lines and the shaded regions indicate the $90\%$ credible regions. The dashed gray lines show the $90\%$ credible bounds of the B-Spline models' prior predictive distributions. The subpopulation distributions are not weighted by their respective branching ratios.}
        \label{fig:spin_distributions}
    \end{centering}
    \script{spin_distributions_plot.py}
\end{figure*}

Figure \ref{fig:spin_distributions} shows the inferred spin magnitude and tilt distributions of \popA{} and \popB{}. We see that the events categorized in \popA{} prefer lower spins and have a stronger preference for alignment than those in \popB{}. Specifically, in the \base{} the tilt distribution of \popA{} peaks at $\cos{\theta}=$ \result{$\CIPlusMinus{\macros[CosTilt][Base][PeakA][max]}$} and has a fraction of negative tilts $f_{\cos{\theta} < 0} = $ \result{$\CIPlusMinus{\macros[CosTilt][Base][PeakA][negfrac]}$} while \popB{}'s spin tilt distribution peaks at $\cos{\theta}=$ \result{$\CIPlusMinus{\macros[CosTilt][Base][ContinuumB][max]}$}, with $f_{\cos{\theta} < 0} = $ \result{$\CIPlusMinus{\macros[CosTilt][Base][ContinuumB][negfrac]}$}. The characteristics of this population -- low mass, low spin magnitudes, and preferential alignment -- are broadly consistent with predictions from isolated binary formation. Due to the large uncertainties in the measured spin parameters, it's important to note that these spin inferences only hint at unique features of a given subset of events. To confidently associate these BBH's with the field formation channel, more observations are needed.

\begin{figure}[]
  \begin{centering}
      \includegraphics[width=\linewidth]{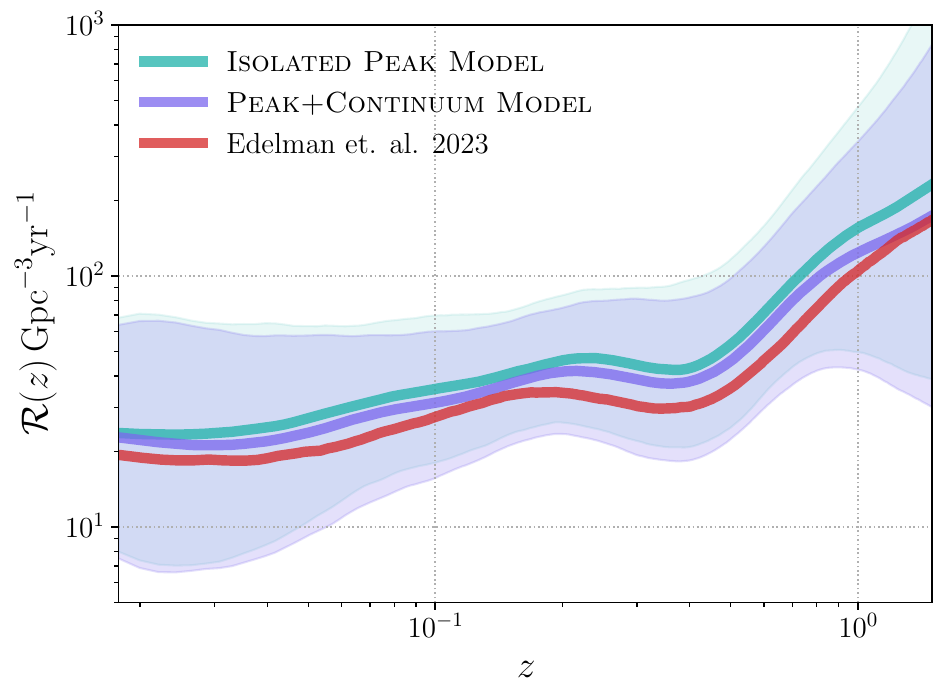}
      \caption{The BBH merger rate as a function of redshift inferred by the \base{}, \comp{}, and \brucepaper. The median curve is shown as a solid line and the shaded regions indicate the $90\%$ credible regions.}
      \label{fig:redshift_distribution}
  \end{centering}
  \script{redshift_plot.py}
\end{figure}

Finally, Figure \ref{fig:redshift_distribution} shows the redshift distributions inferred by the \base{} and \comp{} plotted alongside the distribution inferred by \brucepaper. We assume the redshift distribution is the same for each subpopulation and find both distributions inferred in this work are statistically consistent with \brucepaper.

\begin{figure}[b]
  \begin{centering}
      \includegraphics[width=\linewidth]{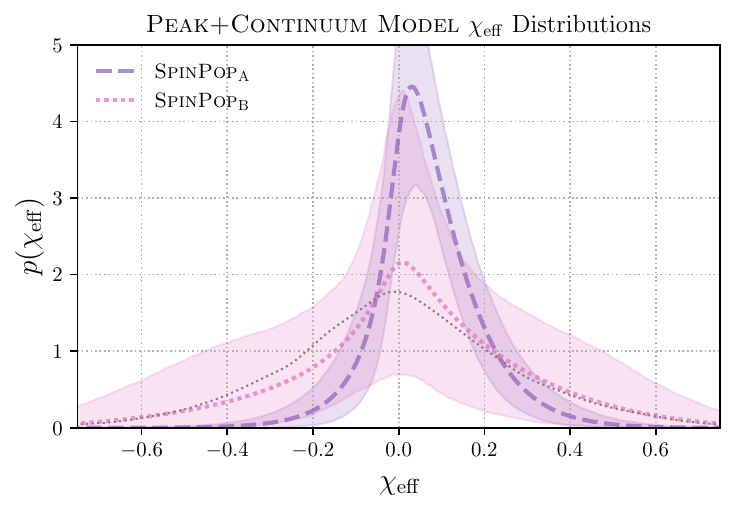}
      \caption{The effective spin distribution inferred by the \comp{}. The medians of the subpopulation component distributions are shown in dashed lines and the shaded regions indicate the $90\%$ credible regions. The subpopulation distributions are not weighted by their respective branching ratios.}
      \label{fig:chi_eff_distributions}
  \end{centering}
  \script{chi_eff_plot.py}
\end{figure}

The primary mass distributions of the \comp{} are shown in the bottom panel of Figure \ref{fig:g1_mass_distribution}, again plotted alongside the total distribution inferred by \brucepaper{}. In this figure, we see that the events in the $\sim15-50 \msun$ range are described by \contA{}, which is the component that shares spin properties with \first{}. The tail end of the mass spectrum is then picked up by \popB{}. Looking to Figure \ref{fig:spin_distributions}, the top right panel shows that the spin magnitude distribution of \popA{} resembles that of \popA{} from the \base{}, while the distribution of \popB{} is completely uninformed. In the bottom right panel, the tilt distribution of \popA{} in the \comp{} also shares similarities with that of \popA{} in the \base{}, and again \popB{} possess an uninformed distribution. Figure \ref{fig:chi_eff_distributions} shows the effective spin distributions of the \comp{}, where we see the \popA{} distribution exhibits an effective spin distribution peaking at positive values while the \popB{} effective spin distribution looks more isotropic, and is symmetric about $\chi_\mathrm{eff} = 0$. Our findings are consistent with previous studies \citep{2111.03634,2110.13542} that have found that the spin magnitudes of BBHs with primary masses $\gtrsim 45-50 \msun$ are more poorly constrained than lower mass binaries, with a tendency toward larger spin magnitudes.

\begin{figure*}[ht!]
  \begin{centering}
      \includegraphics[width=\linewidth]{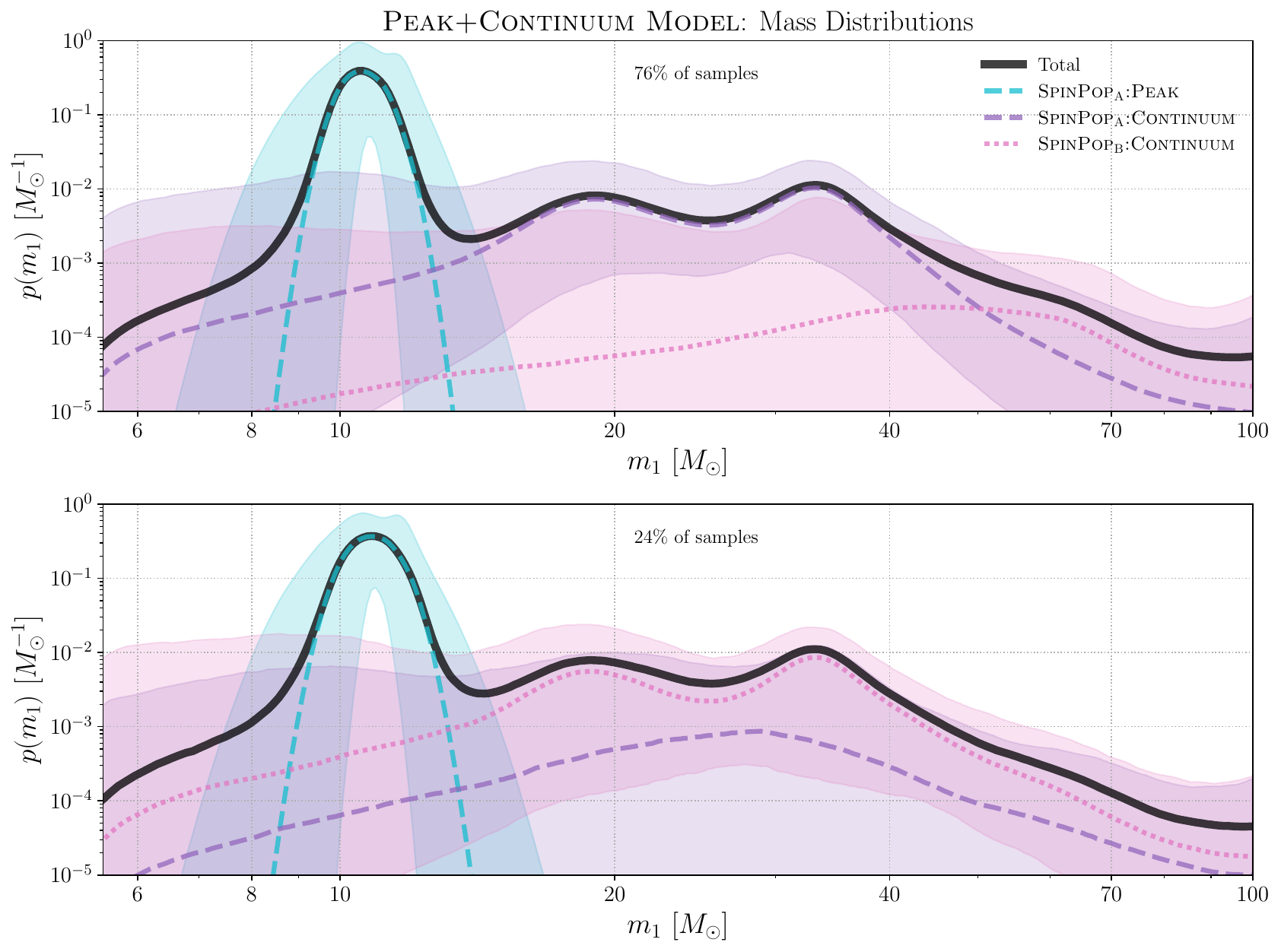}
      \caption{The astrophysical primary mass distributions inferred by the \comp{}. The top (bottom) panel represents $79\%$ ($21\%$) of the posterior draws of \comp{}, specifically posterior draws where the astrophysical branching fraction of \contA{} is greater (lesser) than \contB{}. In both panels, the median total distribution is shown in black. The median of the subpopulation component distributions are shown in dashed lines and the shaded regions indicate the $90\%$ credible regioins. The subpopulation distributions are weighted by their respective branching ratios.}
      \label{fig:g2_mass_distribution}
  \end{centering}
  \script{mass_distribution_g2_plot.py}
\end{figure*}

At first glance, it appears that the events with primary mass $m_1 \sim 15-50 \msun$ are categorized in \popA{} and therefore share the same formation mechanism as the events in the $10\msun$ peak; however, a closer look shows there is some uncertainty in how these events are categorized. To demonstrate this, we plot the primary mass posterior draws inferred when the branching fraction of \contA{} is greater than that of \contB{}, shown in the top panel of Figure \ref{fig:g2_mass_distribution}, and those inferred when the branching fraction of \contA{} is less than \contB{}, shown in the bottom panel of Figure \ref{fig:g2_mass_distribution}. The former makes up $\macros[FracCut][BgreaterA]\%$ of the total samples drawn while the latter makes up $\macros[FracCut][BlessA]\%$, indicating that while the mid-mass events are most likely categorized in \popA{}, there is a small but non-zero probability that they are categorized with the isotropic spin distribution of \popB{}. This could hint that multiple formation channels are responsible for the events in this mass range, though uncertaities are too large to make any confident claims. If this is the case, it is possible that the mass distributions created by different formation mechanisms in this mass range are similar in shape, leaving it up to the spin characteristics to drive event categorization. With spin measurements being generally more uncertain than mass measurements, especially at higher masses, this may lead to the uncertainty we see in the event categorization. 

The location of \first{} in the mass spectrum is robust against our choice of model. We implemented a number of different parametric subpopulations alongside \first{} in our preliminary investigations and still found the location of \first{} at $~10\msun$ in primary mass. Ultimately, we opted for fully data-driven models for the mass distributions of \contA{} and \popB{} and all spin distributions in order to minimize the potential for model misspecification in our results. A recent work, \citet{2303.02973}, conducted a similar study that inferred the existence of two subpopulations in the GWTC-3 catalog data, one with masses $\lesssim 40\msun$ and low spin magnitude and the other with masses in the range $20-90\msun$, isotropic spins, and a spin magnitude distribution peaking at $a_\text{peak} \sim 0.8$. While some of their results are broadly consistent with our results, we do not find evidence for a highly spinning subpopulation and \first{} is the only subpopulation we can claim is distinctly present in the data. Given this, we believe model misspecification may be responsible for their results. In our preliminary investigations using parametric spin models, such as truncated Gaussians, we recovered similar spin distributions as their high-spin group (HSG). We noticed that features in our recovered distributions, specifically the peak in the spin magnitude distribution at $\sim 0.8$, were significantly impacted by our choice of prior boundaries on the standard deviation hyper-parameter.

\subsection{Astrophysical Branching Ratios and Event Categorization}

\begin{figure*}[]
    \begin{centering}
        \includegraphics[width=\linewidth]{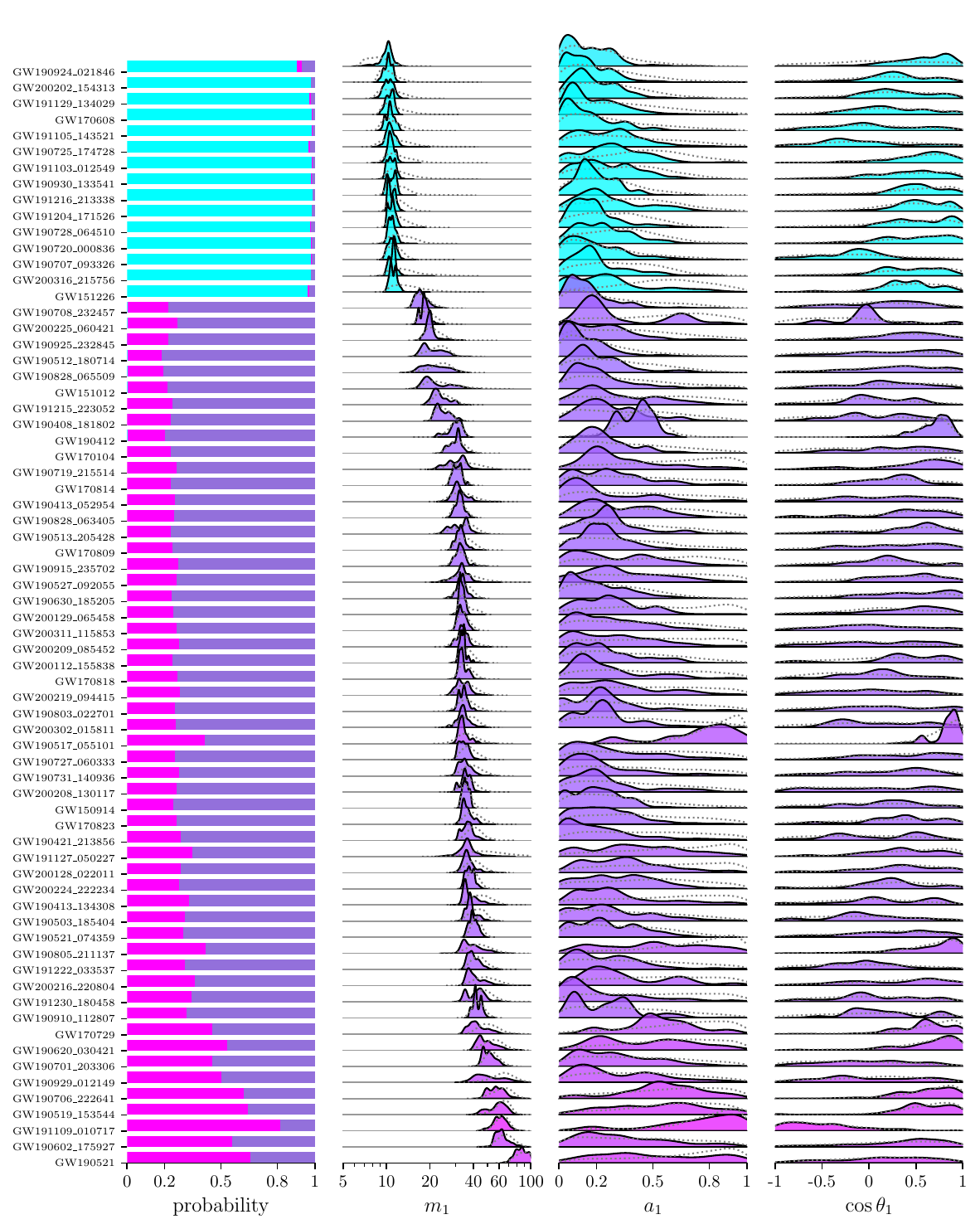}
        \caption{The left most panel shows probability of each event belonging to \first{} (cyan), \contA{} (purple), and \contB{} (magenta). The right three panels show the population reweighed single event primary mass, spin magnitude, and spin tilt posteriors. Gray dashed lines indicate the original unweighed posteriors.}
        \label{fig:ridgeplot}
    \end{centering}
    \script{ridgeplot_marginalized.py}
\end{figure*}

Table \ref{tab:table} lists the astrophysical branching ratios and the number of events constraining each subpopulation/component of the \base{} and \comp{}. The branching ratio and number of events of \first{} are consistent between the \base{} and \comp{}. Figure \ref{fig:ridgeplot} provides a visual representation of event categorization for the \comp{} as well as the physical properties of the subpopulations. Within \contA{}, events GW190412 and GW190517\_055101 are likely outliers of the subpopulation due to their visually different spin distributions. This may indicate that while our models can identify subpopulation-level features in the catalog, categorization of an individual event to a particular subpopulation does not guarantee it is truly a member of that subpopulation. GW190412 and GW190517\_055101 were both detections with fairly certain spin properties. GW190412 was the first clearly unequal mass binary detection, which allowed for a measure of definitively non-zero primary spin magnitude \citep{10.3847/2041-8213/aba8ef, 2010.14527}. GW190517\_055101 was highlighted in the GWTC-2 catalog paper \citet{2010.14527} for having the highest $\chi_\text{eff}$ values, which can be seen in Figure 10 of \citet{2010.14527}. The categorization of these potential outliers may have an impact on the resulting subpopulation distributions, so in future work it will be important to incorporate a method for categorizing outliers that don't fit into any of the given subpopulations. A look at the highest mass events in Figure \ref{fig:ridgeplot} reveals that most of them are characterized with about equal posterior odds between \contA{} and \contB{}.The spin distributions of these events are largely uninformed, so it is not surprising that their categorization is uncertain.

\begin{table*}[]
  \centering
  \begin{tabular}{lcccccc}
  \hline
  \multicolumn{1}{|c|}{\base{}} &
    \multicolumn{1}{c|}{$\lambda$} &
    \multicolumn{1}{c|}{$N_\text{events}$} &
    \multicolumn{1}{c|}{$a_\text{peak}$} &
    \multicolumn{1}{c|}{$\cos{}(\theta)_\text{peak}$} &
    \multicolumn{1}{c|}{$\cos{}(\theta)_{10\%}$} &
    \multicolumn{1}{c|}{$\chi_\text{eff,peak}$} \\ \hline\hline
  \multicolumn{1}{|l|}{\first{}} &
    \multicolumn{1}{c|}{$\CIPlusMinus{\macros[BranchingRatios][Base][PeakA][Frac]}$} &
    \multicolumn{1}{c|}{$\CIPlusMinus{\macros[NumEvents][Base][PeakA]}$} &
    \multicolumn{1}{c|}{$\CIPlusMinus{\macros[SpinMag][Base][PeakA][max]}$} &
    \multicolumn{1}{c|}{$\CIPlusMinus{\macros[CosTilt][Base][PeakA][max]}$} &
    \multicolumn{1}{c|}{$\CIPlusMinus{\macros[CosTilt][Base][PeakA][negfrac]}$} &
    \multicolumn{1}{c|}{$\CIPlusMinus{\macros[ChiEff][Base][PeakA][max]}$} \\ \hline
  \multicolumn{1}{|l|}{\contB{}} &
    \multicolumn{1}{c|}{$\CIPlusMinus{\macros[BranchingRatios][Base][ContinuumB][Frac]}$} &
    \multicolumn{1}{c|}{$\CIPlusMinus{\macros[NumEvents][Base][ContinuumB]}$} &
    \multicolumn{1}{c|}{$\CIPlusMinus{\macros[SpinMag][Base][ContinuumB][max]}$} &
    \multicolumn{1}{c|}{$\CIPlusMinus{\macros[CosTilt][Base][ContinuumB][max]}$} &
    \multicolumn{1}{c|}{$\CIPlusMinus{\macros[CosTilt][Base][ContinuumB][negfrac]}$} &
    \multicolumn{1}{c|}{$\CIPlusMinus{\macros[ChiEff][Base][ContinuumB][max]}$} \\ \hline
  \multicolumn{7}{l}{} \\ \hline
  \multicolumn{1}{|c|}{\comp{}} &
    \multicolumn{1}{c|}{$\lambda$} &
    \multicolumn{1}{c|}{$N_\text{events}$} &
    \multicolumn{1}{c|}{$a_\text{peak}$} &
    \multicolumn{1}{c|}{$\cos{}(\theta)_\text{peak}$} &
    \multicolumn{1}{c|}{$\cos{}(\theta)_{10\%}$} &
    \multicolumn{1}{c|}{$\chi_\text{eff,peak}$} \\ \hline\hline
  \multicolumn{1}{|l|}{\first{}} &
    \multicolumn{1}{c|}{$\CIPlusMinus{\macros[BranchingRatios][Composite][PeakA][Frac]}$} &
    \multicolumn{1}{c|}{$\CIPlusMinus{\macros[NumEvents][Composite][PeakA]}$} &
    \multicolumn{1}{c|}{\multirow{2}{*}{$\CIPlusMinus{\macros[SpinMag][Composite][PeakAContinuumA][max]}$}} &
    \multicolumn{1}{c|}{\multirow{2}{*}{$\CIPlusMinus{\macros[CosTilt][Composite][PeakAContinuumA][max]}$}} &
    \multicolumn{1}{c|}{\multirow{2}{*}{$\CIPlusMinus{\macros[CosTilt][Composite][PeakAContinuumA][negfrac]}$}} &
    \multicolumn{1}{c|}{\multirow{2}{*}{$\CIPlusMinus{\macros[ChiEff][Composite][PeakAContinuumA][max]}$}} \\ \cline{1-3}
  \multicolumn{1}{|l|}{\contA{}} &
    \multicolumn{1}{c|}{$\CIPlusMinus{\macros[BranchingRatios][Composite][ContinuumA][Frac]}$} &
    \multicolumn{1}{c|}{$\CIPlusMinus{\macros[NumEvents][Composite][ContinuumA]}$} &
    \multicolumn{1}{c|}{} &
    \multicolumn{1}{c|}{} &
    \multicolumn{1}{c|}{} &
    \multicolumn{1}{c|}{} \\ \hline
  \multicolumn{1}{|l|}{\contB{}} &
    \multicolumn{1}{c|}{$\CIPlusMinus{\macros[BranchingRatios][Composite][ContinuumB][Frac]}$} &
    \multicolumn{1}{c|}{$\CIPlusMinus{\macros[NumEvents][Composite][ContinuumB]}$} &
    \multicolumn{1}{c|}{$\CIPlusMinus{\macros[SpinMag][Composite][ContinuumB][max]}$} &
    \multicolumn{1}{c|}{$\CIPlusMinus{\macros[CosTilt][Composite][ContinuumB][max]}$} &
    \multicolumn{1}{c|}{$\CIPlusMinus{\macros[CosTilt][Composite][ContinuumB][negfrac]}$} &
    \multicolumn{1}{c|}{$\CIPlusMinus{\macros[ChiEff][Composite][ContinuumB][max]}$} \\ \hline
  \end{tabular}
  \caption{The astrophysical branching ratio $\lambda$ of each subpopulation, the number of events that constrain each subpopulation $N_\text{event}$, and a summary of their spin distributions.}
  \label{tab:table}
  \end{table*}

\subsection{Model Comparison}

We can compute Bayes factors without the need for computing marginal likelihoods using Savage-Dickey Density (SDD) ratios \citep{10.1371/journal.pone.0059655, 10.29220/CSAM.2019.26.2.217}, shown in Table~\ref{tab:BF}. To compute a SDD ratio for two models, one model must be nested within the other as a sharp point hypothesis. In the case of this work, the CYB model is the nested point hypothesis of the \base{} when the branching fraction of \first{}, $\lambda_\textsc{A:P}$, is zero and of the \comp{} when the branching fraction of \first{} and \contA{}, $\lambda_\textsc{A:P}$ and $\lambda_\textsc{A:C}$, are zero. In other words, when the branching fraction of \contB{}, $\lambda_\textsc{B:C}$, in each model is equal to unity. The SDD ratio then gives the Bayes Factor as the ratio between the marginal posterior density of the encompassing model and its prior density evaluated at the null point:

\begin{equation} \label{eq:BF1}
\text{BF}_{\text{CYB,IP}} = \frac{p(\lambda_\textsc{A:P} = 0 | \theta, M_\text{IP})}{p(\lambda_\textsc{A:P} = 0 | M_\text{IP})},
\end{equation}

\begin{equation}\label{eq:BF2}
  \text{BF}_{\text{CYB,P+C}} = \frac{p(\lambda_\textsc{A:P} = \lambda_\textsc{A:C}= 0 | \theta, M_\text{P+C})}{p(\lambda_\textsc{A:P} = \lambda_\textsc{A:P} = 0 | M_\text{P+C})}.
\end{equation}

Since the \base{} is not a point hypothesis of \comp{}, to get $\text{BF}_\text{IP,P+C}$, we take the ratio of the above two Bayes factors:

\begin{equation}\label{eq:BF1}
  \text{BF}_\text{IP,P+C} = \frac{\text{BF}_{\text{CYB,P+C}}}{\text{BF}_{\text{CYB,IP}}}.
\end{equation}

One draw back of the SDD ratio is its sensitivity to undersampling. We therefore ensured we sufficiently sampled the tails of the $\lambda$ distributions in each model by collecting a large number of samples ($\sim\mathcal{O}(10^6)$). 
As seen in Table \ref{tab:BF}, both the \base{} and \comp{} are strongly preferred over the CYB model. This is likely due both to the support in the data for the $10\msun$ peak having a unique spin distribution, as well as the sharpness of the $10 \msun$ peak requiring a characteristically smaller  (log)mass scale than the other features in the spectrum. The preference for the \comp{} over the \base{} shows that the data support the hypothesis that a non-zero fraction of the binaries found elsewhere in the mass spectrum have spin characteristics consistent with those in the $10 \msun$ peak, while others might have spins more consistent with the highest mass events. How the events outside the $10 \msun$ peak are categorized will likely become more clear when this analysis is performed on the LVK's next GWTC release, which will include data collected during the LVK's fourth observing run.



\section{Astrophysical Interpretation} \label{sec:astro}


Spin predictions of binaries formed in the stellar field are dependent on the many physical processes that may occur prior to the stellar binary becoming a BBH.  Spin magnitude of a BH can depend sensitively on the efficiency of angular momentum (AM) transport between its progenitor stellar core and envelope \citep{2203.02515}. Efficient AM transport, such as through the Taylor-Spruit magnetic dynamo \citep{10.1051/0004-6361:20011465}, leads to low spinning BHs \citep{10.3847/2041-8213/ab339b} while less efficient AM transport, such as that predicted by the shellular model \citep{1992A&A...265..115Z,2012A&A...537A.146E,10.3847/1538-4365/aacb24,2019MNRAS.485.4641C}, can preserve the spin of the progenitor star. However, the effects of AM transport by these mechanisms can be obfuscated by tidal interactions between the binary components, accretion, and mass transfer, which can spin up the binary or increase AM transport efficiency, thus shedding spin. Natal supernova kicks are thought to be the leading cause of spin orbit misalignments in field binaries, while tidal forces and mass transfer tend to align BH spins with the orbital AM. As discussed in \citet{2111.03634}, recent studies of BBH formation in globular clusters predict \emph{suppressed} merger rates for binaries with primary BH masses $\lesssim 10 \msun$~\citep{2009.01861,1906.10260,1808.04514}.  Therefore, BBH formation in globular clusters is unlikely to be responsible for the BBHs in the $10\msun$ peak, and is most likely a subdominant formation mechanism in the catalog.

While we find that the $10\msun$ subpopulation is consistent with the general characteristics associated with field formation, large uncertainties in both spin measurements and predictions from population synthesis models prevent us from placing informative constraints on the formation physics. If the $10\msun$ subpopulation is indeed a product of isolated binary evolution, then our inferred spin distributions hint at this channel producing binaries with low, modestly misaligned spins. This could indicate that AM transport is efficient in massive stars, as modeled by variations of the Taylor-Spruit magnetic dynamo \citep{1706.07053}, and that natal kicks are a common occurrence during field BBH formation.

Our results are consistent with other analyses that find the data does not require a discontinuous non-spinning subpopulation \citep{arXiv2205.08574,2205.12329, 2209.02206, 2301.01312,10.48550/arXiv.2302.07289, 2210.12287}, though is in tension with other analyses that have claimed its existence \citep{doi.org/10.3847/2041-8213/ac2f3c,2105.10580}; however, we cannot completely rule out a non-spinning population. The $20\msun$ and $35\msun$ peaks may also be consistent with field formation, as we infer spin properties similar to those of the $10\msun$ peak; though, as we discussed in Section \ref{sec:results} there is a possibility that some events in this mass range are more consistent with field formation spin characteristics. The astrophysical branching ratios we infer in the \comp{} are consistent with the field and dynamical branching ratios inferred by \citet{2011.10057}, $[\beta_{\text{field}}, \beta_{\text{dynamical}}] = [0.86^{+0.11}_{-0.36}, 0.14^{0.36}_{-0.11}]$, under the assumption that \popA{} is consistent with field formation and \contB{} is more consistent with dynamical formation.

The sharp fall-off in primary mass of \contA{} in the \comp{} could give an estimate of the lower edge of the PISN mass gap. The 99th percentile of the primary mass distribution of \contA{} is $m_{1,99\%} = $ \result{$\CIPlusMinus{\macros[Mass][Composite][ContinuumA][99percentile]}$ \msun}, which is consistent with predictions that place the lower edge of the gap between $40-70\msun$ \citep{1901.00215,1910.12874v1,2103.07933v1,2104.07783v2}.

\section{Conclusion} \label{sec:conclusion}

\begin{table*}[ht!]
    \centering
    \begin{tabular}{@{}cccc@{}}
    \toprule
    Model & CYB & $\base{}$ & $\comp{}$ \\ \midrule
    CYB & 0 & $-\macros[LogBayesFactors][IP_to_CYB]$ & $-\macros[LogBayesFactors][PC_to_CYB]$ \\
    $\base{}$ & \macros[LogBayesFactors][IP_to_CYB] & 0 & $-\macros[LogBayesFactors][PC_to_IP]$ \\
    $\comp{}$ & \macros[LogBayesFactors][PC_to_CYB] & \macros[LogBayesFactors][PC_to_IP] & 0 \\
    \bottomrule
    \end{tabular}
    \caption{$\log_{10}$ Bayes factors of the \base{}, \comp{}, and the Cover Your Basis (CYB) model (\brucepaper). The values follow the format $\log_{10} \text{BF}_\text{row,col}.$; e.g. the \comp{} row and CYB column shows the $\log_{10}$ Bayes factor of the \comp{} relative to the CYB model.}
    \label{tab:BF}
    \end{table*}

As the catalog of compact object mergers continues to grow, we are able to probe the physical properties of these systems with higher fidelity and uncover details in their distributions previously obscured by our lack of data. With these advancements comes the ability to piece together formation histories imprinted in the details of these distributions. Understanding the physical properties of compact binaries and their formation has implications for the broader astrophysics community such as providing constraints on stellar evolution theories and population synthesis simulations, the physics of globular clusters, the impact of stellar metallicity, neutron star equations of state, and much more.

By leveraging the hierarchical Bayesian inference toolkit, a mixture of parametric and data-driven models, and combining information across mass and spin, we were able to identify a peak in the BBH primary mass spectrum at $m_\text{1,peak} = $ \result{$\CIPlusMinus{\macros[Mass][Base][PeakA][max]}$ \msun} that corresponds to a subpopulation of BBH's with low spins and a preference for alignment, consistent with isolated binary formation. We then extended our \base{} to search the rest of the mass spectrum for events with similar spin characteristics to the $10\msun$ subpopulation. We found that the $20\msun$ and $35\msun$ peaks are consistent with the $10\msun$ events, though there \textit{may} be an underlying subpopulation consistent with the highest mass events. We find evidence of a distinct population at high masses whose spin properties are largely uninformed by the current catalog.

Due to the large uncertainties that are currently present in both population synthesis models and the measurements of BBH properties, we are unable to place strong constraints on the physics behind isolated binary evolution or (P)PISN. However, if the $10\msun$, $20\msun$, and $35\msun$ subpopulations are truly products of these channels, they likely produce binaries with low spins, though whether these formation channels are dominated by non-spinning BHs remains unclear. Aligned spin systems are also not completely ruled out, but the subpopulations appear to possess modest misalignments and the apparent fall-off of the $35\msun$ peak may indicate a lower bound on the PISN mass gap of \result{$\CIPlusMinus{\macros[Mass][Composite][ContinuumA][99percentile]}$ \msun}.

The data driven models used in this analysis and developed in the python library \textsc{GWInferno} can inform us about the full compact binary catalog beyond identifying BBH subpopulations. Currently, the LVK categorizes mergers as binary black holes, binary neutron stars, or neutron star binary black holes \emph{a priori} based on mass thresholds and tidal deformation values. Instead of categorizing merger components \emph{a priori} and then fitting the mass and spin distributions of each category individually, discrete latent variables could be used to simultaneously classify merger components and infer their mass and spin distributions.

In future work, we will incorporate a method for classifying outliers to subpopulations as well as testing this framework on a simulated catalog in order to better understand how uninformed data affects our results. Repeating this analysis on the data from the next observing run -- which is expected to substantially increase the current catalog size -- should provide more insights with stronger constraints on subpopulation properties.

\section{Acknowledgements}\label{sec:acknowledments}
This material is based upon work supported by the National Science Foundation Graduate Research Fellowship under Grant No. 2236419. This research has made use of data, software and/or web tools obtained from the Gravitational Wave Open Science Center 
(\url{https://www.gw-openscience.org/}), a service of LIGO Laboratory, the LIGO Scientific Collaboration and the Virgo Collaboration. 
The authors are grateful for computational resources provided by the LIGO Laboratory and supported by National Science Foundation Grants PHY-0757058 and PHY-0823459.  
This work benefited from access to the University of Oregon high performance computer, Talapas. This material is based upon work supported 
in part by the National Science Foundation under Grant PHY-2146528 and work supported by NSF's LIGO Laboratory which is a major facility 
fully funded by the National Science Foundation.
\software{
\textsc{Showyourwork}~\citep{2110.06271},
\textsc{Astropy}~\citep{1307.6212, 1801.02634, 2206.14220},
\textsc{NumPy}~\citep{10.1038/s41586-020-2649-2},
\textsc{SciPy}~\citep{10.1038/s41592-019-0686-2},
\textsc{Matplotlib}~\citep{10.1109/MCSE.2007.55},
\textsc{Jax}~\citep{github.com/google/jax},
\textsc{NumPyro}~\citep{1810.09538,1912.11554},
}
\bibliography{bib}{}
\bibliographystyle{aasjournal}


\appendix
\section{Model Priors} \label{sec:priors}

\begin{table*}[h!]
    \centering
    \begin{tabular}{|l|l|l|l|}
    \hline
    \textbf{Model} & \textbf{Parameter} & \textbf{Description} & \textbf{Prior} \\ \hline \hline
    \multicolumn{4}{|c|}{\textbf{Primary Mass Model Parameters}} \\ \hline

    \first & $\mu_m$ & Mean of Truncated Log Peak & $ \sim \mathrm{N}_{LT}(\ln10,\ln5, \text{a}=5) $ \\ \cline{2-4} 
    & $\sigma_m$ & Standard Deviation of Truncated Log Peak & $\sim \mathrm{N}_T(0, 0.2, \text{a}=0.01, \text{b}=0.3)$ \\ \hline

    \contA & $\bm{c}$ & Basis coefficients & $\sim \mathrm{Smooth}(\tau_\lambda, \sigma, r, n)$ \\ \cline{2-4} 
     and & $\tau_\lambda$ & Smoothing Prior Scale & 1 \\ \cline{2-4}
     \contB & $r$ & order of the difference matrix for the smoothing prior & 1 \\ \cline{2-4} 
     & $\sigma$ & width of Gaussian priors on coefficients in smoothing prior & 15 \\ \cline{2-4} 
     & $n$ & number of basis functions in the basis spline & 48 \\ \hline \hline

    \multicolumn{4}{|c|}{\textbf{Mass Ratio Model Parameters}} \\ \hline
    All Mass Ratio & $\bm{c}$ & Basis coefficients & $\sim \mathrm{Smooth}(\tau_\lambda, \sigma, r, n)$ \\ \cline{2-4} 
    Models & $\tau_\lambda$ & Smoothing Prior Scale & 1 \\ \cline{2-4}
    & $r$ & order of the difference matrix for the smoothing prior & 1 \\ \cline{2-4} 
    & $\sigma$ & width of Gaussian priors on coefficients in smoothing prior & 5 \\ \cline{2-4} 
    & $n$ & number of basis functions in the basis spline & 30 \\ \hline \hline

    \multicolumn{4}{|c|}{\textbf{Redshift Evolution Model Parameters}} \\ \hline
    All Redshift Models & $\lambda$ & slope of redshift evolution power law $(1+z)^\lambda$ &  $\sim \mathcal{N}(0,3)$ \\ \cline{2-4}
    & $\bm{c}$ & Basis coefficients & $\sim \mathrm{Smooth}(\tau_\lambda, \sigma, r, n)$ \\ \cline{2-4} 
     & $\tau_\lambda$ & Smoothing Prior Scale & 1 \\ \cline{2-4}
     & $r$ & order of the difference matrix for the smoothing prior & 2 \\ \cline{2-4} 
     & $\sigma$ & width of Gaussian priors on coefficients in smoothing prior & 10 \\ \cline{2-4} 
     & $n$ & number of knots in the basis spline & 20 \\ \hline \hline

    \multicolumn{4}{|c|}{\textbf{Spin Distribution Model Parameters}} \\ \hline
    All Spin Magnitude & $\bm{c}$ &  Basis coefficients & $\sim \mathrm{Smooth}(\tau_\lambda, \sigma, r, n)$  \\ \cline{2-4} 
    Models & $\tau_\lambda$ & Smoothing Prior Scale & 25\\ \cline{2-4}
    & $r$ & order of the difference matrix for the smoothing prior & 2 \\ \cline{2-4} 
    & $\sigma$ & width of Gaussian priors on coefficients in smoothing prior & 5 \\ \cline{2-4} 
    & $n$ & number of knots in the basis spline & 16 \\ \hline \hline

    All Tilt Models & $\bm{c}$ &  Basis coefficients & $\sim \mathrm{Smooth}(\tau_\lambda, \sigma, r, n)$  \\ \cline{2-4} 
    & $\tau_\lambda$ & Smoothing Prior Scale & 25 \\ \cline{2-4}
    & $r$ & order of the difference matrix for the smoothing prior & 2 \\ \cline{2-4} 
    & $\sigma$ & width of Gaussian priors on coefficients in smoothing prior & 5 \\ \cline{2-4} 
    & $n$ & number of knots in the basis spline & 16 \\ \hline \hline
    \end{tabular}
    \caption{All hyperparameter prior choices for each of the basis spline models from this manuscript. See \brucepaper{} for more detailed description of basis spline or smoothing prior parameters.}
    \label{tab:model_priors}
\end{table*}

\section{Reproducibility}
\label{sec:reproducibility}

This study was carried out using the reproducibility software \href{https://github.com/showyourwork/showyourwork}{\showyourwork} \citep{2110.06271}, which leverages continuous integration to programmatically download the data from \href{https://zenodo.org/}{zenodo.org}, create the figures, and compile the manuscript. Each figure caption contains two links: one to the dataset stored on zenodo used in the corresponding figure, and the other to the script used to make the figure (at the commit corresponding to the current build of the manuscript). The git repository associated to this study is publicly available at \url{https://github.com/jaxengodfrey/CosmicCousins}. The datasets are stored at \href{https://zenodo.org/}{zenodo.org} at \url{https://zenodo.org/records/10892979}

\end{document}